This is the postprint of the article published in Solar Energy Materials and Solar Cells, Volume 200, 15 September 2019, 110022. https://doi.org/10.1016/j.solmat.2019.110022

Data Sheets are available at https://doi.org/10.5281/zenodo.3335833

# Effect of the organic cation on the optical properties of Lead Iodine perovskites

César Tablero Crespo
*Instituto de Energía Solar, E.T.S.I. de Telecomunicación,*
*Universidad Politécnica de Madrid,*
*Ciudad Universitaria s/n, 28040 Madrid, SPAIN.*
*e-mail: ctablero@etsit.upm.es*

**Abstract**

The effect of 14 organic cations on the optical properties of the lead iodine perovskite is analyzed. The electronic and optical properties are obtained using first principles. The absorption coefficients are split into inter-atomic species components in order to quantify all of the contributions. For energies close to the bandgaps, the main contribution is from the Pb-Pb intra-species transitions. For higher energy this contribution is still important in addition to I-I and Pb-I contributions, and to the 3 and 4 species term. Almost all absorption properties are qualitatively similar. Furthermore, this absorption coefficient splitting also allows the optical characteristics that the substitution of Pb by another element should satisfy to be identified in order to reduce the toxicity because of Pb while maintaining a high absorption capacity.

## 1. Introduction.

Lead iodine perovskites $APbI_3$ are promising semiconductors for solar energy conversion. The most studied so far is the methylammonium $CH_3NH_3PbI_3$ because of its electronic properties, low non-radiative recombination, and its high device efficiencies15-20 % [1,2]. In addition to their efficiencies, the perovskites have attracted interest because of the novel device structures and new perovskite materials [3–7].

Data Sheets are available at https://doi.org/10.5281/zenodo.3335833

The electronic properties of perovskite compounds are determined mainly by the $PbI_6$ octahedra (Figure 1). Although cation A does not directly contribute to the electronic properties, its size can cause distortion in the Pb-I distances and angles. Therefore, indirectly, the A cations affect the electronic properties. Higher efficiency is still possible through structural modification using different organic cations A.

Furthermore, the instability of the lead iodine perovskites, in particular due to water degradation, could be suppressed by exchanging the organic cation A. There are a lot of possibilities for the replacement of the organic cation. In the literature the influence of the cation on the electronic structure has been studied for some cations ( $[C_6H_5C_2H_4NH_3]_2^+$ [8], formamidinium [9]). In this work we have analyzed the electronic and optical properties of 14 organic cations as detailed in Table 1.

One of the properties that determine the conversion of solar energy is the absorption coefficient. For this reason we will focus our study on this optical property by evaluating its variation with different organic cations. We will use first principles to evaluate the electronic and optical properties. In addition to analyzing the absorption coefficients in depth, we will split them into contributions due to excitations between the atomic species that make up the material. This decomposition allows us to identify and quantify the most important contributions as a function of the photon energy. This will be described in the methodology section. Later the results are presented and compared with other theoretical and experimental results. Finally, the main conclusions of this work are set out.

## 2. Methodology

The electronic and optical properties are obtained from first-principles calculations based on density-functional theory (DFT) [10,11]. We use the Perdew-Burke-Ernzerhof

Data Sheets are available at https://doi.org/10.5281/zenodo.3335833

[12] exchange-correlation functional. Core electrons are dealt with by Troullier–Martins [13] pseudopotentials expressed in the Kleinman–Bylander [14,15] form. Wave functions are expanded in a polarized double-zeta basis set made up of localized pseudoatomic orbitals [16]. For all cations the calculations have been carried out with spin-polarization, periodic boundary conditions and 256 (405) special k points in the irreducible Brillouin zone for the electronic (optical) properties. The structure parameters used in the calculation were taken from reference [17].

The imaginary part of the dielectric function is calculated within the random phase approximation according to:

$$e_2(E) \sim C\frac{1}{E^2}\sum_{\lambda}\sum_{\mu>\lambda}\int d\mathbf{k}(f_\lambda - f_\mu)\,|\,p_{\mu\lambda}\,|^2\,\delta(E_\mu - E_\lambda - E) \qquad (1)$$

where $\lambda$ and $\mu$ label the bands of the system, $p_{\lambda\mu}$ is the momentum matrix element between $\lambda$ and bands, and $E_\lambda$ and $f_\lambda$ are the corresponding energies and occupations of the $\lambda$-bands. The other optical properties were obtained from the imaginary part of the dielectric function using the Kramers-Kronig relationships [18].

We have developed a method [19] for calculating the contributions to the optical properties starting from first-principles calculations. Firstly the $p_{\lambda\mu}$ are split into different species contributions in accordance with $p_{\mu\lambda} = \sum_A \sum_B p_{\mu\lambda}^{AB}$. When A≠B, $p_{\mu\lambda}^{AB}$ is an inter-specie element that couples the basis set functions on different species atoms A and B. If A=B, $p_{\mu\lambda}^{AA}$ represents an intra-species contribution. Since the optical properties depend on the square of the momentum operator, they can be expressed as a many-specie expansion; for example, the absorption coefficient $\alpha = A_{12} + A_{34}$. The first term involves intra-species (A=B) and inter-species (A≠B) contributions $A_{12} = \sum_A \sum_B \alpha_{AB}$. $A_{34}$ involves 3-

Data Sheets are available at https://doi.org/10.5281/zenodo.3335833

species (A≠B≠C) and 4-species (A≠B≠C≠D) contributions. Note that there are no terms larger than those of 4 species. Obviously the terms of two, three and four species will only be present if the compound has more than one, two, and three species respectively. Usually $A_{34}$ is much smaller than $A_{12}$ and $\alpha \approx A_{12}$, except in compounds with a large number of species. This is the case of the perovskites analyzed in this work.

## 3. Results and discussion

Figure 2a shows the values of the bandgap obtained for all the organic cations analyzed (Table 1). To compare, the results from reference [17] with different exchange-correlation functional (GGA from Perdew-Wang (GGA-PW) [20] and Heyd-Scuseria-Ernzerhof (HSE06) [21]) and basis set (plane wave basis set), are also shown. The bandgap results of this work are in general larger than those obtained using the GGA-PW and lower than those obtained using the more sophisticated, and expensive computationally, HSE06. In accordance with other theoretical analyses in the literature [22], generally the use of a localized basis set combined with the GGA-PBE functional provides reasonable estimates of the bandgaps and cell geometry in organic-inorganic metal halide perovskites. It represents a compromise between accuracy and computational cost. Among the reasons for this good match with the experimental results may be: (i) with respect to the basis set, a localized basis set is able to represent better the limited combination of states between the organic cations and the $PbI_6$ octahedron than a plane wave basis set for energies close to the bandgap; (ii) with respect to the GGA methodology, it is widely known that the GGA usually underestimates the bandgap due to the self-correlation errors. On the other hand, because of the presence of Pb, the relativistic effects such as spin−orbit coupling, which is not considered here, would

Data Sheets are available at https://doi.org/10.5281/zenodo.3335833

reduce the band gap. Therefore, a reasons for the good agreement between GGA and experimental results can be the error compensation. More sophisticated methods with relativistic effects and with plane wave basis sets could compare worse despite being more accurate.

The bandgap comparison among the organic cations in Table 1 with the most studied A=Me cation (Figure 2b) indicate that the largest difference is around +0.31 eV for A=$Hy_1$. This not only happens with the exchange-correlation functional used in this work but with other exchange-correlation functional and basis sets (Figure 2b). In addition to the bandgap, as we will show later, this cation also has notable differences in the absorption coefficients with respect to the rest of the organic cations analyzed. This larger difference with respect to other cations is because the A=$Hy_1$($[(OH)NH_3]^+$) is the only one that contains oxygen but not carbon. The bandgap differences (Figure 2b) compare well with some experimental results in the literature. For example, the Formamidinium (Fo) experimental bandgap decreases 0.02-0.07 eV [9,23] with respect to Methylammonium (Me), similar to that obtained in this work (Figure 2b).

For these perovskites the valence band (VB) edge states consist mainly of p(I) states, whereas the conduction band (CB) edge states come from the p(Pb) states with a smaller contribution from the p(I) states. The electronic levels of the organic cation atoms lie deep within the VB and CB. It indicates that their contribution to the electronic properties near to bandgap is very small. Then the band edges electronic properties are directly related to the $PbI_6$ octahedra. The organic cations interact weakly with the inorganic $PbI_6$ octahedra and their electron orbitals have little combination with the

Data Sheets are available at https://doi.org/10.5281/zenodo.3335833

iodine and lead states. However, the size of the organic cation can distort the $PbI_6$ octahedra, and therefore, indirectly influence the electronic properties.

For the absorption of solar radiation and its subsequent energy conversion, the most important optical property is the absorption coefficient (AC). The ACs obtained for the lead iodine perovskites $APbI_3$ with all the A organic cations in Table 1 are shown in Figure 3. Almost all ACs are roughly similar (Figure 3) except for the A=$Hy_1$ organic cation. As has already been mentioned this cation has the largest bandgap difference compared with the most studied A=Me cation. The shape of the AC-$Hy_1$ is similar to the other cations but displaced towards higher energies as a result of the larger bandgap. In the Ref. [25], Figures 3-7 corresponding to the total and split AC are represented individually for each organic cation.

The AC results reproduce the experimental large absorption band between 3.2-3.5 eV [24] for A=Me, and the decrease between 3.5 and 4 eV. Also the peak around 3.10 eV that appears in the experimental absorption spectrum [25] of the perovskite with A=Me is reproduced by the results of this work (Figure 3). These peaks are also observed for almost all organic cations. Therefore, these results, according to Figure 3, are quite general, except for the aforementioned A=$Hy_1$.

The ACs in Figure 3 do not provide information on the different contributions of excitations between the different states of the species. In order to extract this information and thus be able to analyze these contributions, the ACs have been split into species contributions. The more important intra- and inter-species contributions are the Pb-Pb (Figure 4) and I-I (Figure 5) intra-species contributions, and the inter-species Pb-I (Figure

This is the postprint of the article published in Solar Energy Materials and Solar Cells, Volume 200, 15 September 2019, 110022. https://doi.org/10.1016/j.solmat.2019.110022

Data Sheets are available at https://doi.org/10.5281/zenodo.3335833

6). The other intra- and inter-species contributions that involve C, H and N are negligible compared to the previous ones.

Furthermore, as the number of species of all perovskites studied is larger than or equal to 4, the terms involving three and four species ( $A_{34}$) could be important. This does not happen in the most common binary semiconductors (II-VI, II-VI, etc) where the 3 and 4 species terms are absent. Almost all the perovskites analyzed in this work, except that one with the organic cation A= $Hy_2$, have five different species: Pb, I, C (O for the A= $Hy_1$ cation), N, and H. The number of one, two, three, and four different species terms in the split of the absorption coefficients are 5, 10, 10, and 5 respectively. Then the number of 3 and 4 species terms is similar to the number of one and two inter-species terms. Note that for the single specie semiconductors (A=Si, Ge, etc) the many-body species expansion has only one (intra-specie) term $\alpha_{A\text{-}A}$ . In the case of binary III-V, II-VI, etc, semiconductors ( $A_xB_y$=GaAs, ZnTe, etc), the expansion has two (intra-specie) terms ( $\alpha_{A\text{-}A}$ and $\alpha_{B\text{-}B}$), and one two species (inter-specie) term ( $\alpha_{A\text{-}B}$). The terms of three (four) species are only present for ternary (quaternary) and higher compounds. In addition, as it was previously mentioned in the methodology section, there are no terms greater than four species in the many-body species expansion of optical properties. Note that this decomposition is exact and serves to evaluate the different contributions to the absorption coefficient. The $A_{34}$ term, represented in Figure 7, also has a high contribution similar to the Pb-Pb, I-I, and Pb-I contributions, but for higher energies. In fact, $A_{34}$ is the highest term between around 3 eV for some cations. The transitions contribute to the optical properties for energies corresponding to the difference between the energies of the final and initial transition states. As it has been aforementioned, the electronic states of the

Data Sheets are available at https://doi.org/10.5281/zenodo.3335833

organic cation atoms lie deep within the VB and CB whereas Pb and I have high contributions for energies close to the energy bandgap. Therefore the contributions of the $A_{34}$ term, which involves at least one organic cation atom, corresponds to higher energies than the Pb-Pb, I-I, and Pb-I contributions. They also have a similar contribution because the number of 3 and 4 species terms is similar to the number of one and two inter-species terms.

The optical absorption for energies close to the bandgap is due to the Pb-Pb intra-specie excitation, mainly to the s(Pb)-p(Pb) contribution, in accordance with previous results [26]. From Figures 4-6 (Pb-Pb, I-I, and Pb-I contributions respectively) the larger intra- and inter-species contributions to the experimental absorption band for A=Me between 3.2-3.5 eV [24] are, in descending order of importance, the I-I, Pb-I and Pb-Pb species transition. An analysis of the states involved indicate that these excitations are due mainly to s(I)-p(I), p(Pb)-p(I) and s(Pb)-p(Pb). This characteristic absorption band in the absorption spectrum of the A=Me cation is very similar to the rest of the organic cations (Figures 4-6). Similarly, the peak around 3.10 eV in the A=Me experimental absorption spectrum [25], present for almost all cations (Figure 3), is due to mainly to I-I (Figure 5) and Pb-I (Figure 6), and with a lower proportion to Pb-Pb. These contributions mainly came from s(I)-p(I), p(Pb)-p(I) and s(Pb)-p(Pb) respectively.

The optical absorption band between 2.5-2.7 eV [24] for A=Me is also mainly due to the intra-specie transition Pb-Pb [26], although with a large I-I contribution (Figure 5). It can be observed more clearly in the Ref. [25] where the total and split AC for each cation are represented separately. This band is present for almost all cations, although the size of the absorption band is different depending on the organic cation.

Data Sheets are available at https://doi.org/10.5281/zenodo.3335833

The more important difference of the A= $Hy_1$ cation with respect to other cations is the displacement of the absorption band between 2.5-2.7 eV for the intra-species term Pb-Pb (Figure 4) towards higher energies (~3.0-3.3 eV).

Both the experimental absorption band between 3.2-3.5 eV [24] and the peak around 3.10 eV also have a high contribution of terms of 3 and 4 species ( $A_{34}$, Figure 7), which in some cases is even higher than the I-I and Pb-I.

## 4. Conclusions

The electronic and optical properties of the 14 organic lead iodine perovskites $APbI_3$ are obtained using DFT first principles. The electronic properties are determined mainly by the Pb-I interactions in the $PbI_6$ octahedra: the edges of the VB and CB are predominantly made up of p(I) and p(Pb) states respectively.

After that, the ACs are obtained, compared with experimental results, and the differences between the organic cations analyzed. Furthermore, the ACs have been split into inter-species components in order to identify and quantify the origin of the peaks in the experimental absorption spectra. For energies close to the bandgap, the main contribution is from the Pb-Pb (s(Pb)-p(Pb)) intra-species transitions. A wide absorption band above 3 eV is present in almost all organic perovskites. It mainly came from of the I-I, Pb-I and Pb-Pb species transition (s(I)-p(I), p(Pb)-p(I) and s(Pb)-p(Pb) respectively). However, the 3 and 4 species term $A_{34}$ has a contribution similar to the previous ones, and even higher in the case of some cations. Therefore, although the organic cations do not directly contribute to the optical properties through intra- and inter-species terms, they do contribute indirectly via terms of 3 and 4 species.

Data Sheets are available at `https://doi.org/10.5281/zenodo.3335833`

Almost all ACs are qualitatively similar, except for the A= $Hy_1$, as a consequence of the larger bandgap. Therefore, except for this cation, the absorption properties do not vary considerably. However, with respect to other properties that determine the final efficiency and stability of the device, the effect of organic cations could be decisive.

The content of lead in these compounds can generate environmental and toxicity problems. To maintain the high absorption capacity and reduce toxicity, the M element that would substitute lead should have a high M-M inter-species contribution.

**Acknowledgments**

This work has been supported by the National Spanish project INVENTA-PV (TEC2015-64189-C3-1-R, MINECO/FEDER) and the European Commission through the funding of the project GRECO (H2020-787289).

Data Sheets are available at `https://doi.org/10.5281/zenodo.3335833`

## Lists of references

[1] J. Burschka, N. Pellet, S.-J. Moon, R. Humphry-Baker, P. Gao, M.K. Nazeeruddin, M. Grätzel, Sequential deposition as a route to high-performance perovskite-sensitized solar cells, Nature. 499 (2013) 316.

[2] M.A. Green, A. Ho-Baillie, H.J. Snaith, The emergence of perovskite solar cells, Nat. Photonics. 8 (2014) 506.

[3] T. Zhou, M. Wang, Z. Zang, X. Tang, L. Fang, Two-dimensional lead-free hybrid halide perovskite using superatom anions with tunable electronic properties, Sol. Energy Mater. Sol. Cells. 191 (2019) 33–38. doi:https://doi.org/10.1016/j.solmat.2018.10.021.

[4] X. Zeng, T. Zhou, C. Leng, Z. Zang, M. Wang, W. Hu, X. Tang, S. Lu, L. Fang, M. Zhou, Performance improvement of perovskite solar cells by employing a CdSe quantum dot/PCBM composite as an electron transport layer, J. Mater. Chem. A. 5 (2017) 17499–17505. doi:10.1039/C7TA00203C.

[5] M. Wang, Z. Zang, B. Yang, X. Hu, K. Sun, L. Sun, Performance improvement of perovskite solar cells through enhanced hole extraction: The role of iodide concentration gradient, Sol. Energy Mater. Sol. Cells. 185 (2018) 117–123. doi:https://doi.org/10.1016/j.solmat.2018.05.025.

[6] T. Zhou, Y. Zhang, M. Wang, Z. Zang, X. Tang, Tunable electronic structures and high efficiency obtained by introducing superalkali and superhalogen into AMX3-type perovskites, J. Power Sources. 429 (2019) 120–126. doi:https://doi.org/10.1016/j.jpowsour.2019.04.111.

[7] B. Yang, M. Wang, X. Hu, T. Zhou, Z. Zang, Highly efficient semitransparent CsPbIBr2 perovskite solar cells via low-temperature processed In2S3 as electron-transport-layer, Nano Energy. 57 (2019) 718–727. doi:https://doi.org/10.1016/j.nanoen.2018.12.097.

[8] J. Gebhardt, Y. Kim, A.M. Rappe, Influence of the Dimensionality and Organic Cation on Crystal and Electronic Structure of Organometallic Halide Perovskites, J. Phys. Chem. C. 121 (2017) 6569–6574. doi:10.1021/acs.jpcc.7b00890.

[9] N.J. Jeon, J.H. Noh, W.S. Yang, Y.C. Kim, S. Ryu, J. Seo, S.I. Seok, Compositional engineering of perovskite materials for high-performance solar cells, Nature. 517 (2015) 476.

[10] W. Kohn, L.J. Sham, Self-Consistent Equations Including Exchange and Correlation Effects, Phys Rev. 140 (1965) A1133–A1138. doi:10.1103/PhysRev.140.A1133.

[11] J.M. Soler, E. Artacho, J.D. Gale, A. García, J. Junquera, P. Ordejón, Daniel Sánchez-Portal, The SIESTA method for ab initio order- N materials simulation, J. Phys. Condens. Matter. 14 (2002) 2745.

[12] J.P. Perdew, K. Burke, M. Ernzerhof, Generalized Gradient Approximation Made Simple, Phys Rev Lett. 77 (1996) 3865–3868. doi:10.1103/PhysRevLett.77.3865.

[13] N. Troullier, J.L. Martins, Efficient pseudopotentials for plane-wave calculations, Phys Rev B. 43 (1991) 1993–2006. doi:10.1103/PhysRevB.43.1993.

[14] L. Kleinman, D.M. Bylander, Efficacious Form for Model Pseudopotentials, Phys Rev Lett. 48 (1982) 1425–1428. doi:10.1103/PhysRevLett.48.1425.

[15] D.M. Bylander, L. Kleinman, 4f resonances with norm-conserving pseudopotentials, Phys Rev B. 41 (1990) 907–912. doi:10.1103/PhysRevB.41.907.

[16] O.F. Sankey, D.J. Niklewski, Ab initio multicenter tight-binding model for molecular-dynamics simulations and other applications in covalent systems, Phys Rev B. 40 (1989) 3979–3995. doi:10.1103/PhysRevB.40.3979.

[17] C. Kim, T.D. Huan, S. Krishnan, R. Ramprasad, A hybrid organic-inorganic perovskite dataset, Sci. Data. 4 (2017) 170057.

[18] G.F. Bassani, Electronic states and optical transitions in solids, Pergamon Press, Oxford,New York, 1975.

[19] C. Tablero, An evaluation of BiFeO3 as a photovoltaic material, Sol. Energy Mater. Sol. Cells. 171 (2017) 161–165. doi:https://doi.org/10.1016/j.solmat.2017.06.049.

[20] É.D. Murray, K. Lee, D.C. Langreth, Investigation of Exchange Energy Density Functional Accuracy for Interacting Molecules, J. Chem. Theory Comput. 5 (2009) 2754–2762. doi:10.1021/ct900365q.

[21] J. Heyd, G.E. Scuseria, M. Ernzerhof, Hybrid functionals based on a screened Coulomb potential, J. Chem. Phys. 118 (2003) 8207–8215. doi:10.1063/1.1564060.

Data Sheets are available at `https://doi.org/10.5281/zenodo.3335833`

[22] N. Hernández-Haro, J. Ortega-Castro, Y.B. Martynov, R.G. Nazmitdinov, A. Frontera, DFT prediction of band gap in organic-inorganic metal halide perovskites: An exchange-correlation functional benchmark study, Chem. Phys. 516 (2019) 225–231. doi:https://doi.org/10.1016/j.chemphys.2018.09.023.

[23] M.T. Weller, O.J. Weber, J.M. Frost, A. Walsh, Cubic Perovskite Structure of Black Formamidinium Lead Iodide, α-[HC(NH2)2]PbI3, at 298 K, J. Phys. Chem. Lett. 6 (2015) 3209–3212. doi:10.1021/acs.jpclett.5b01432.

[24] N. Kitazawa, Y. Watanabe, Y. Nakamura, Optical properties of CH3NH3PbX3 (X = halogen) and their mixed-halide crystals, J. Mater. Sci. 37 (2002) 3585–3587. doi:10.1023/A:1016584519829.

[25] J.-H. Im, C.-R. Lee, J.-W. Lee, S.-W. Park, N.-G. Park, 6.5% efficient perovskite quantum-dot-sensitized solar cell, Nanoscale. 3 (2011) 4088–4093. doi:10.1039/C1NR10867K.

[26] M. Hirasawa, T. Ishihara, T. Goto, Exciton Features in 0-, 2-, and 3-Dimensional Networks of [PbI6]4- Octahedra, J. Phys. Soc. Jpn. 63 (1994) 3870–3879. doi:10.1143/JPSJ.63.3870.

[25] C. Tablero Crespo, *Data in Brief, "submitted".*

Data Sheets are available at https://doi.org/10.5281/zenodo.3335833

**List of Tables**

| Formula | Cation (A) | abbreviation |
|---|---|---|
| $[(CH_3)NH_3]^+$ | Methylammonium | Me |
| $[(CH_3)_2NH_2]^+$ | Dimethylammonium | Di |
| $[(CH_3)_3NH]^+$ | Trimethylammonium | Tr |
| $[(CH_3)_4N]^+$ | Tetramethylammonium | Te |
| $[CH_3CH_2NH_3]^+$ | Ethylammonium | Et |
| $[CH_3(CH_2)_2NH_3]^+$ | Propylammonium | Pr |
| $[(CH_3)_2CHNH_3]^+$ | Isopropylammonium | Is |
| $[CH_3C(NH_2)_2]^+$ | Acetamidinium | Ac |
| $[C_3H_5N_2]^+$ | Imidazolium | Im |
| $[C_3H_8N]^+$ | Azetidinium | Az |
| $[CH(NH_2)_2]^+$ | Formamidinium | Fo |
| $[C(NH_2)_3]^+$ | Guanidinium | Gu |
| $[(OH)NH_3]^+$ | Hydroxylammonium | $Hy_1$ |

Data Sheets are available at `https://doi.org/10.5281/zenodo.3335833`

| | | |
|---|---|---|
| $[NH_3NH_2]^+$ | Hydrazinium | $Hy_2$ |

**Table 1**: Organic cations (A) of the Lead Iodine perovskites $APbI_3$ evaluated in this work.

## List of Figures

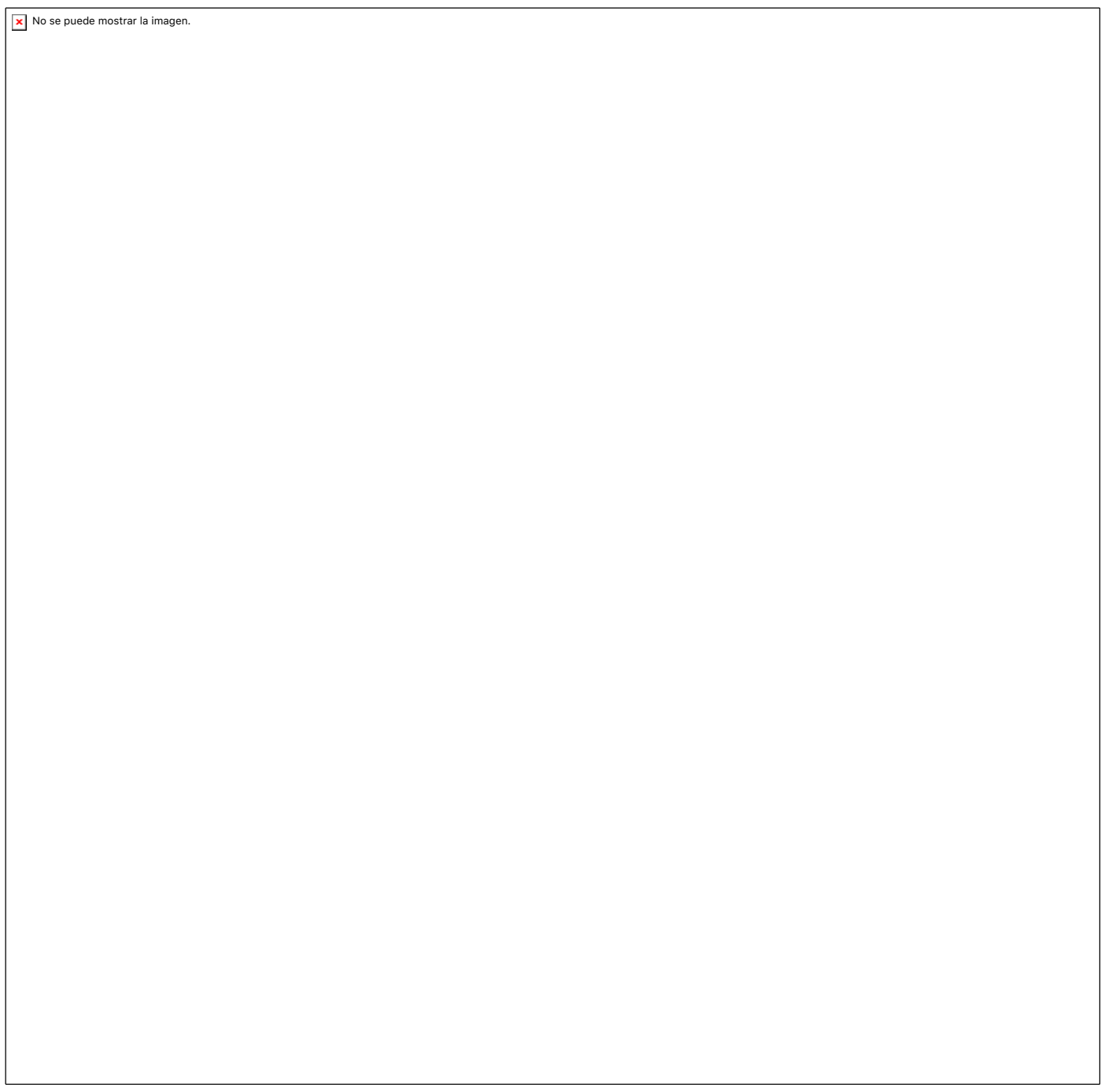

Data Sheets are available at `https://doi.org/10.5281/zenodo.3335833`

**Figure 1**: Crystalline structure of the Perovskite $APbI_3$ semiconductors. The shaded areas show the $PbI_6$ octahedra.

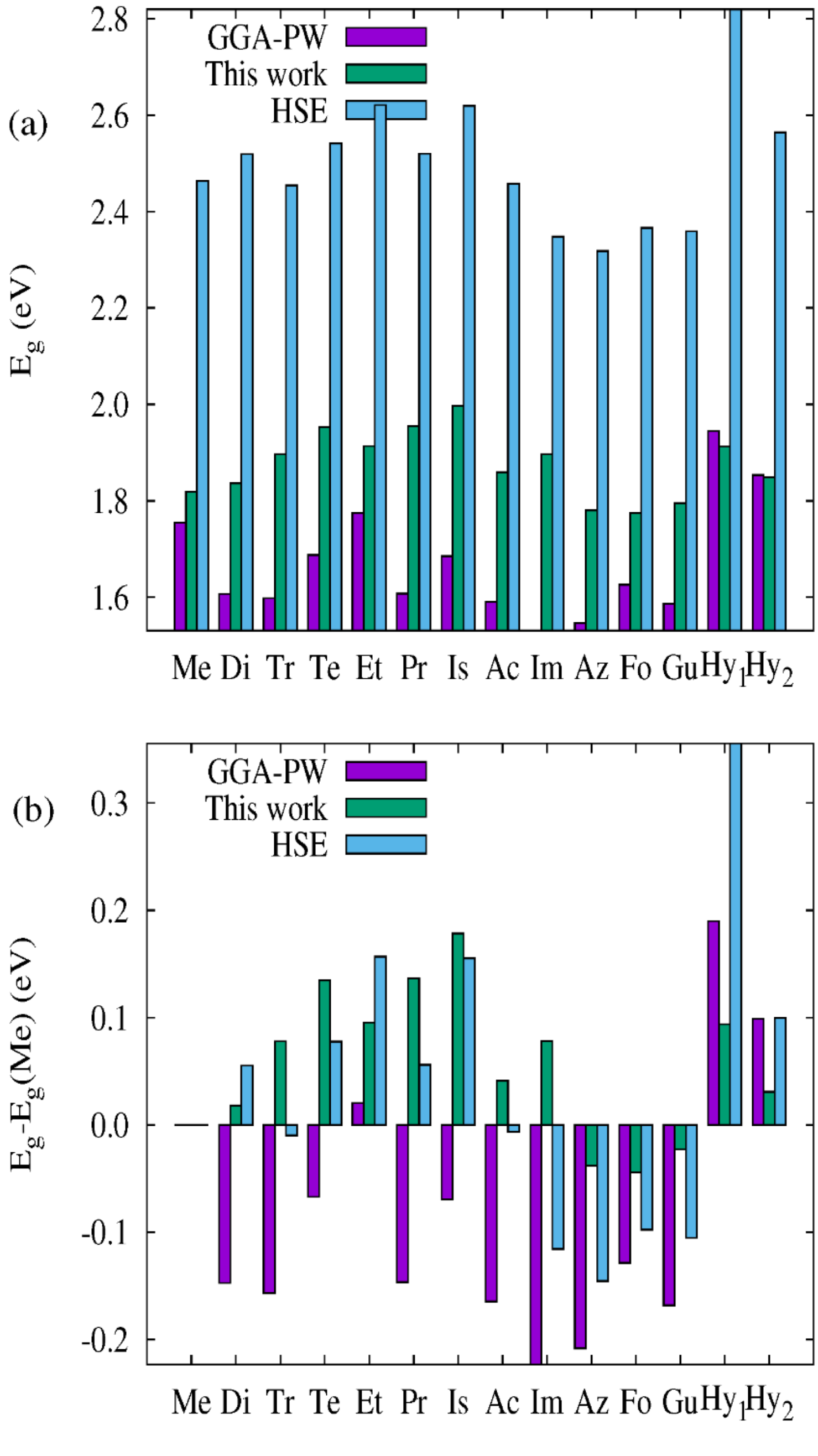

**Figure 2:** (a) Comparison of the energy bandgap ( $E_g$ ) obtained in this work for several organic cations (Table 1) with other results in the literature using different methodology

Data Sheets are available at https://doi.org/10.5281/zenodo.3335833

[17]. (b) Difference between the bandgap with respect to the A=Me bandgap ( $E_g$-$E_g$(Me) ).

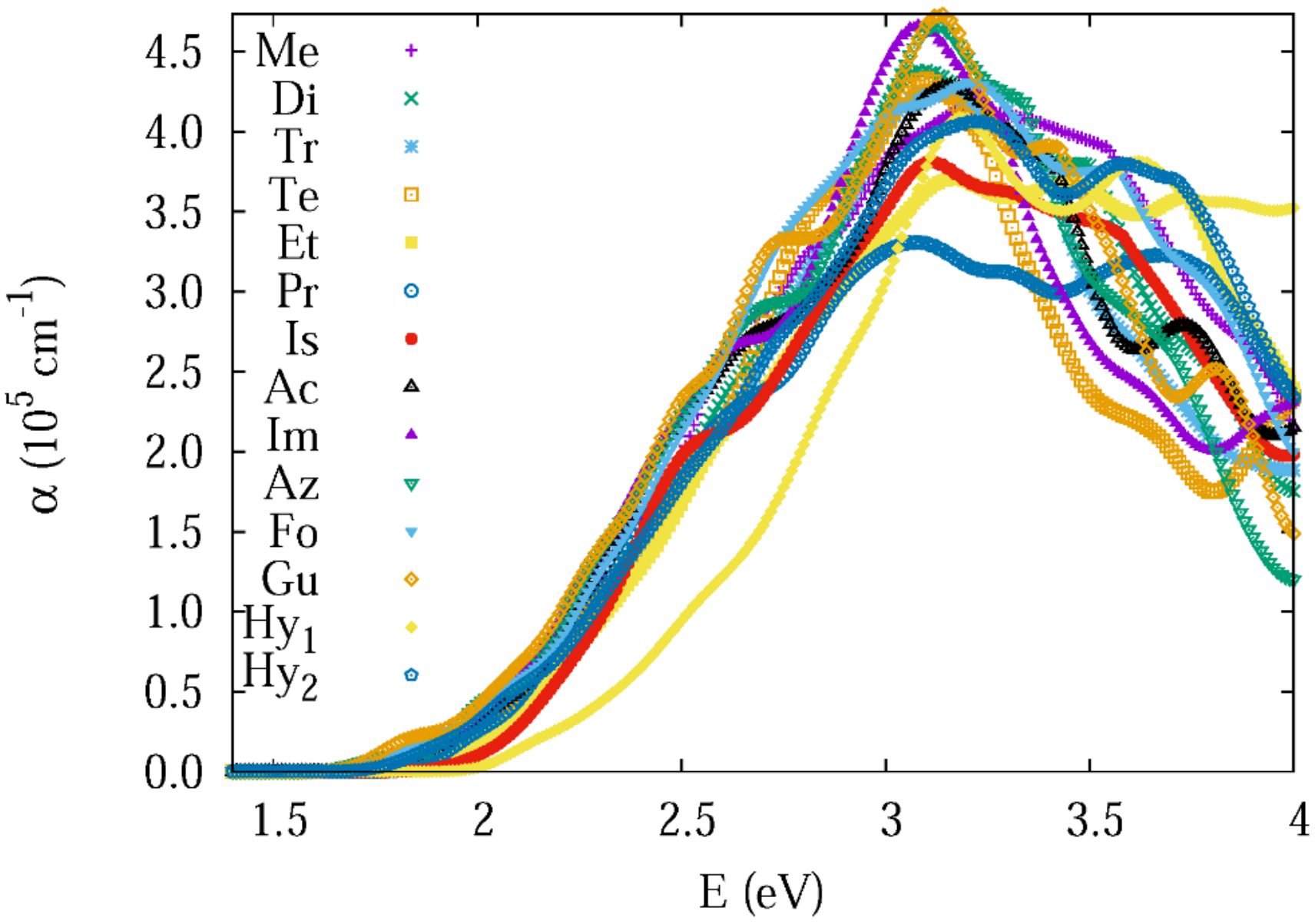

**Figure 3:** $APbI_3$ absorption coefficient for several organic A cations (Table 1).

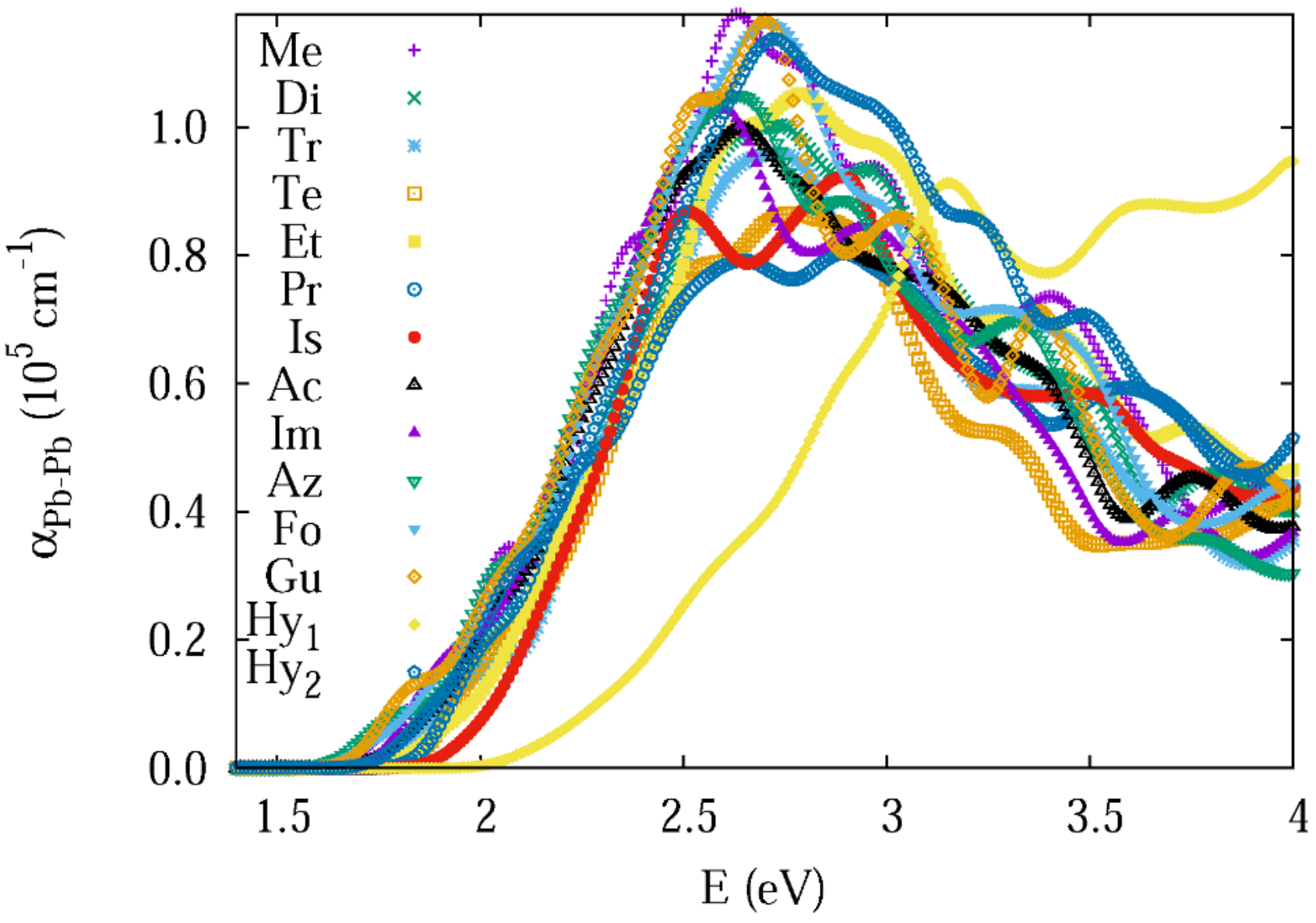

**Figure 4:** Same legend as that of Figure 3, but for the absorption coefficient split into Pb-Pb species contributions ( $\alpha_{Pb\text{-}Pb}$).

Data Sheets are available at `https://doi.org/10.5281/zenodo.3335833`

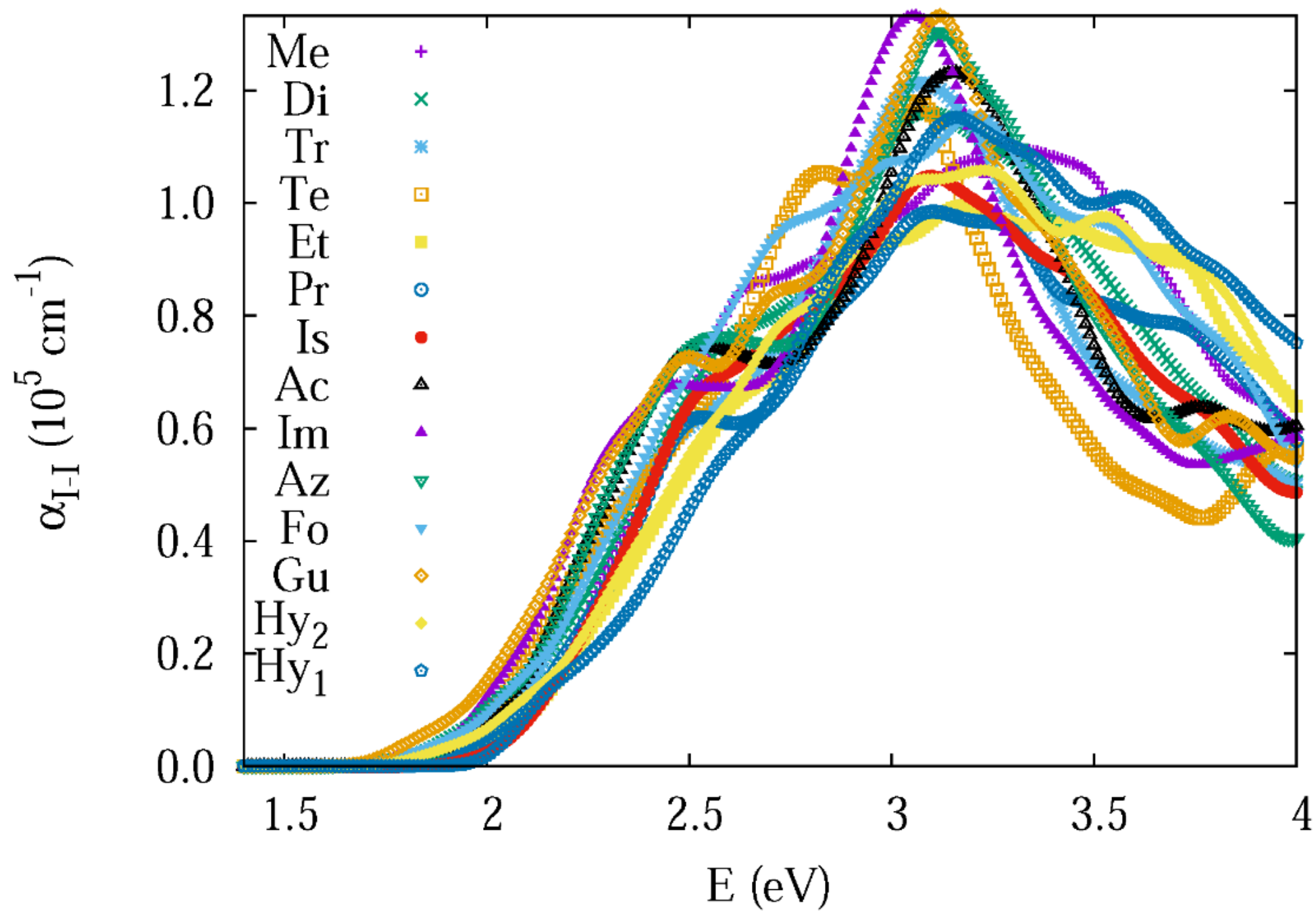

**Figure 5:** Same legend as that of Figure 3, but for the absorption coefficient split into I-I species contributions ($\alpha_{\text{I-I}}$).

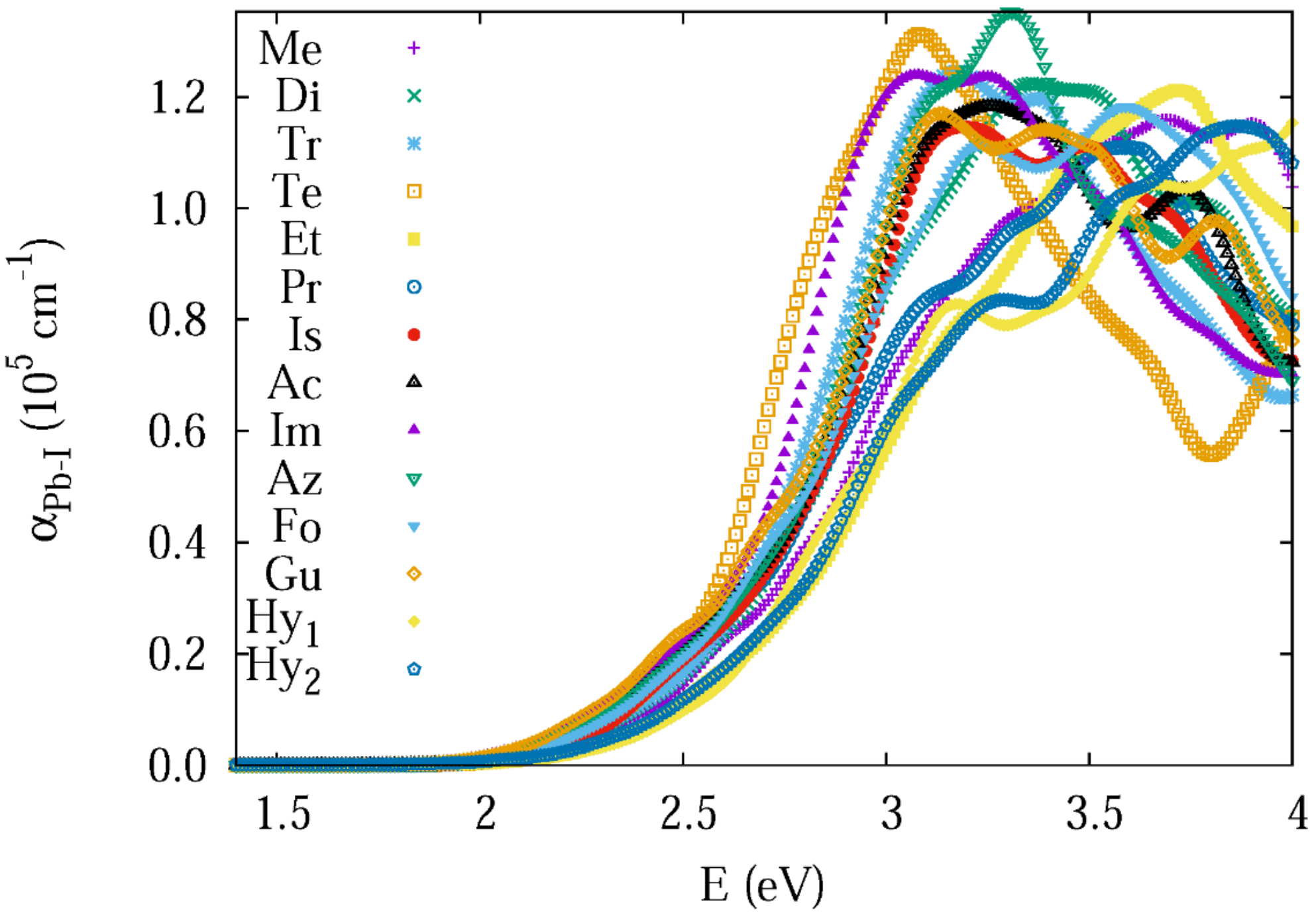

**Figure 6:** Same legend as that of Figure 3, but for the absorption coefficient split into Pb-I species contributions ($\alpha_{\text{Pb-I}}$).

Data Sheets are available at `https://doi.org/10.5281/zenodo.3335833`

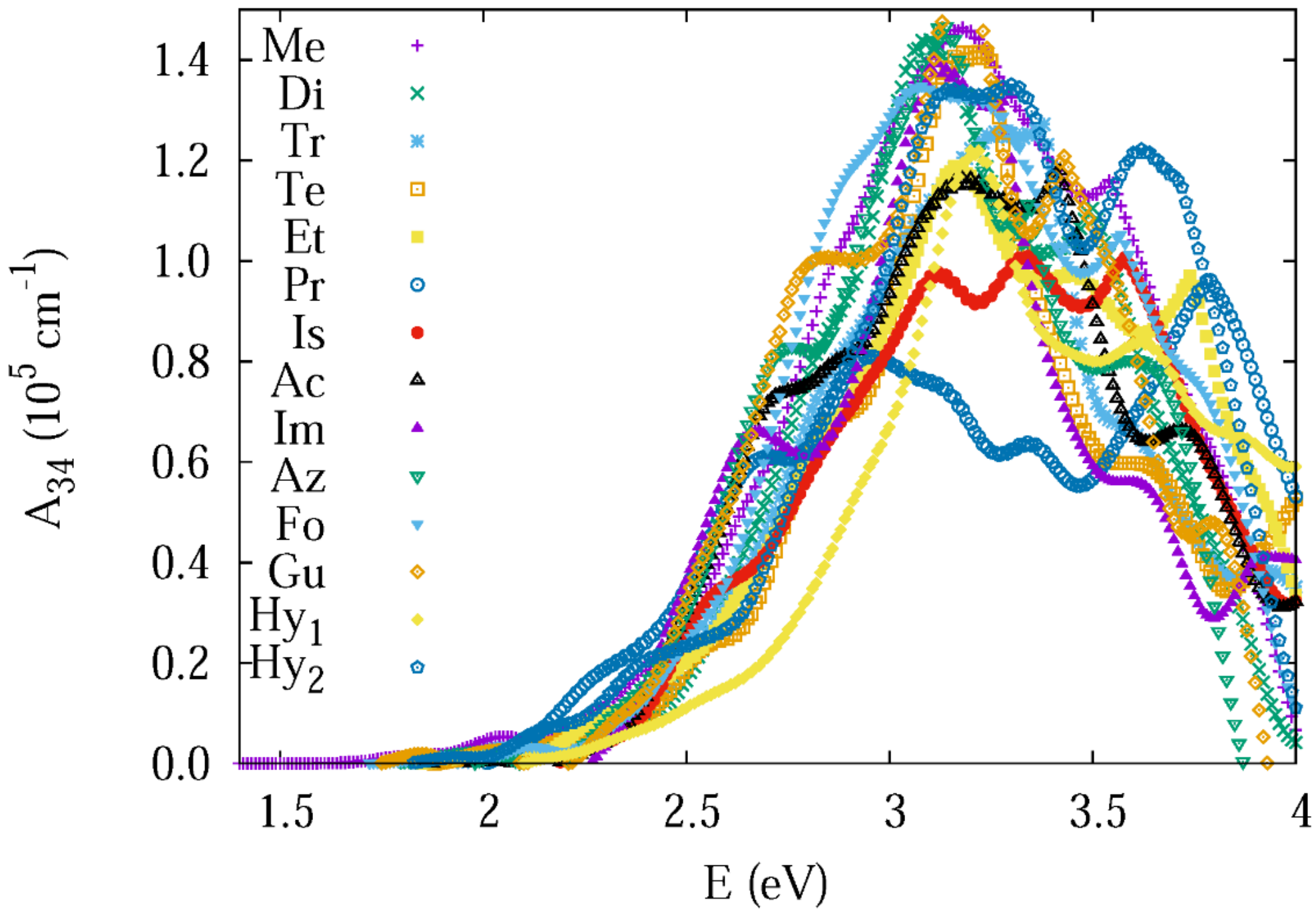

**Figure 7:** Same legend as that of Figure 3, but for the 3- and 4-species contributions ($A_{34}$).